\pdfoutput=1

\documentclass[11pt]{article}

\usepackage{acl}

\usepackage{times}
\usepackage{latexsym}

\usepackage[T1]{fontenc}

\usepackage[utf8]{inputenc}

\usepackage{microtype}
\usepackage{algorithm}
\usepackage{algorithmic}
\usepackage{multirow}
\usepackage{graphicx}
\usepackage{booktabs}
\usepackage{amsmath}
\usepackage{amsfonts} 
\newcommand{\model}{Ruleformer}
\title{Ruleformer: Context-aware Rule Mining over Knowledge Graph}

\author{
	Zezhong Xu$^{1}$ , Peng Ye$^{1}$ ,
	Hui Chen$^{3}$, Meng Zhao$^{4}$\\
	\textbf{
	Huajun Chen$^{1,2}$, Wen Zhang$^{1}$\thanks{$\quad$ Corresponding Author.} 
	} \vspace{1.0mm} \\
	$^1$Zhejiang University \& AZFT Joint Lab for Knowledge Engine, China \\
	$^2$Hangzhou Innovation Center, Zhejiang University\\
	$^3$Alibaba Group, $^4$Huawei \\
	\fontsize{11}{10}\selectfont 
	\{xuzezhong, yep, huajunsir, zhang.wen\}@zju.edu.cn,  \\
	\fontsize{11}{10}\selectfont weidu.ch@alibaba-inc.com, zhaomeng57@huawei.com \\
}

\begin{document}
\maketitle
\begin{abstract}
Rule mining is an effective approach for reasoning over knowledge graph (KG). Existing works mainly concentrate on mining rules. However, there might be several rules that could be applied for reasoning for one relation, and how to select appropriate rules for completion of different triples has not been discussed. In this paper, we propose to take the context information into consideration, which helps select suitable rules for the inference tasks. Based on this idea, we propose a transformer-based rule mining approach, \textbf{\model}\footnote{Source code of Ruleformer is available at https://github.com/zjukg/ruleformer.}. It consists of two blocks: 
1) an encoder extracting the context information from subgraph of head entities with modified attention mechanism, 
and 2) a decoder which aggregates the subgraph information from the encoder output and generates the probability of relations for each step of reasoning.
The basic idea behind {\model} is regarding rule mining process as a sequence to sequence task. To make the subgraph a sequence input to the encoder and retain the graph structure, we devise a relational attention mechanism in Transformer. The experiment results show the necessity of considering these information in rule mining task and the effectiveness of our model.
\end{abstract}

\section{Introduction}

People built different Knowledge Graphs, such as Freebase~\cite{bollacker2008freebase}, to store complex structured information and knowledge about abstract and real world. The facts in KG are usually represented in the form of triplets, e.g., \textit{(New York, isCityOf, USA)}. 
KGs have been widely applied in various intelligent systems such as question answering \cite{DBLP:conf/naacl/YasunagaRBLL21,DBLP:conf/sigir/Chen0ZZYXC22}, recommender system\cite{guo2020survey, DBLP:conf/icde/WongFZVCZHCZC21, DBLP:conf/icde/ZhangWYWZC21} and zero-shot learning\cite{DBLP:conf/kdd/GengCZXCPHXC22, DBLP:conf/semweb/0007CGPYC21}.

Although these KGs already contain a large number of relations and entities, they still suffer from the incompleteness of facts, whether constructed automatically or manually. In order to further expand KGs, many reasoning methods have been proposed to automated fact exploration, such as knowledge graph embedding (KGE)~\cite{TransE,ComplEx,ConvE,RotatE,DBLP:conf/wsdm/ZhangPZBC19}, graph neural networks~\cite{RGCN}, and rule mining methods~\cite{IEEE2018Rudik,galarraga2013amie,yang2017differentiable,sadeghian2019drum,DBLP:conf/www/ZhangPWCZZBC19}. 
Compared with deep learning approaches like KGE, rule mining methods are preferred due to their interpretability for reasoning and robustness for domain knowledge transfer.
To mine the structure and confidence of rules at the same time in a fast way, differentiable rule learning methods\cite{yang2017differentiable,sadeghian2019drum} are introduced and attract many research interests.
 
The learning targets of existing methods is determining the confidence and structure of rules, according to which reasoning tasks are conducted. However, the uniform way of applying rules ignores the context of specific triplets, for which rules should be applied in different order.
 
\begin{figure}[tb]
\centering
  \includegraphics[scale=0.6]{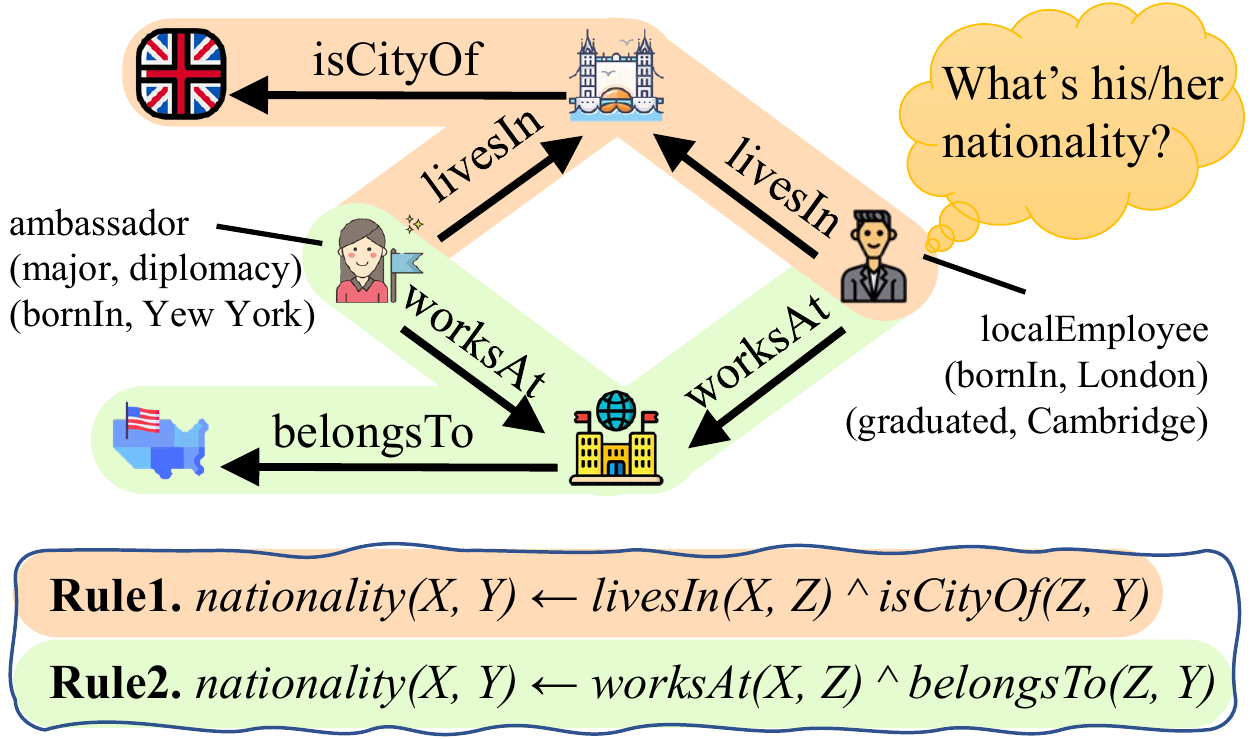}
  \caption{Rules in different context.}
  \label{fig:example}
\end{figure}

Specifically, for a query $(h,r,?)$, there might be multiple rules of relation $r$ (not only one) that could be used to get the target answer. For a specific query triplet, not all rules of $r$ are convincing for reasoning, but how to select appropriate rule has not been studied.
For example, to infer the missing fact, \textit{(}\texttt{localEmployee/ambassador}\textit{, nationality, ?)} as Figure \ref{fig:example} illustrated, previous rule mining methods only rely on the head relation \textit{nationality}, so rules of \textit{nationality} will be used in a fixed order. The two queries will get the same results, and whatever the order is, one of answers will be wrong. More precious rules can be observed if we pay attention to the head entity's identity, which is \texttt{localEmployee/ambassador}. 
Based on the observation, when the head entity is a \texttt{localEmployee}, the rule 
 \textit{nationality(X, Y)  $\gets$ livesIn(X, Z) $\land$ isCityOf(Z, Y)} 
 may derive more accurate result. In another case, if we know the head entity is an \texttt{ambassador}, then rule 
 \textit{nationality(X, Y)  $\gets$ worksAt(X, Z) $\land$ belongsTo(Z, Y)} 
 will be a better choice for inferring. 
This example shows that the order and choice of rules to be applied for reasoning will significantly affect the results, and context information could help determine proper order.
 
In this paper, we investigate making rule mining methods not only learn confidence and structures of rules but also learn to choose suitable rules based on the context of the head entity in completion tasks, which is challenging because of the diversity of context. 
We propose a Transformer-based model to aggregate the context around the head entity because of its excellent performance on information interaction.
Essentially, the model consists of two blocks, 1) an encoder block aggregating the information of the head entity's subgraph and completing the information intersection of context, 2) a decoder block utilizing the aggregated entity embedding to generate the probability of relations for each step in differentiable rule mining process.
Since Transformer framework is a sequence to sequence model, we design a converter that can turn graph structure into sequence. Moreover, to maintain the information of subgraph, we modify the attention mechanism of Transformer by adding the relation information to the attention calculation process, which we call relational attention mechanism.
 
Experimentally, we evaluate our model on several datasets (\textit{UMLS}~\cite{umls}, \textit{FB15K-237}~\cite{fb15k237}, \textit{WN18RR}~\cite{ConvE}) on link prediction and rule parsing task. The improvement in both experiments demonstrate the effectiveness of {\model}. A case study is also analyzed, and the case proves our assumption and the ability of {\model} to select suitable rules for different triples. 
 
In summary, our contributions are as follows:
\begin{itemize}
\item We draw attention to the problem of mining and applying suitable rules depending on the specific context for reasoning task in KG.

\item We propose a new model, {\model}, that can aggregate the information of subgraph and use the context to support the reasoning process.
\item The experiment results prove that our model outperforms existing rule mining methods on link prediction task and rule quality assessment. It successfully selects suitable rules according to the exploitation of context. 
\end{itemize}

\section{Related work}

\subsection{Rule Learning Methods}
The problem of learning rules over KG can be seen as a type of statistical relational learning~\cite{koller2007introduction}.
AMIE~\cite{galarraga2013amie} concentrates on association rule mining   with three operations, including dangling atom, instantiated atom and closing atom that add different type of atoms to incomplete rules and uses pre-defined evaluation metrics to prune incorrect rules. AMIE+~\cite{galarraga2015fast} revises the rule extending process and improves evaluation method based on AMIE.

Anyburl~\cite{AnyBURL} proposed an framework that can mine rules in an effcient way. Based on the randomly sampled path, it replace some entities with variables and get rules. 

Rudik~\cite{IEEE2018Rudik} can mines positive and negative rules. The positive rules can be used to infers new facts in KG, and the negative rules are useful for other tasks, like detecting erroneous triples.

Generally, conventional symbolic-based rule learning methods are built on effective search strategy, pruning techniques and pre-defined static evaluation indicators. The inference processes are transparent while they may suffer from large search space.

More recently, differentiable rule learning methods based on TensorLog~\cite{cohen2016tensorlog} are proposed, which can learn the confidence and structure of rule at the same time. Neural-LP~\cite{yang2017differentiable} use RNN to generate the possibilities of different relation for each step, and the parameters can be optimized in a differentiable way.
Based on Neural-LP, DRUM~\cite{sadeghian2019drum} is proposed, which use low rank approximation to get better results.
Neural-Num-LP~\cite{2020_Num} extends Neural-LP to learn the numerical rules.
Neural Logic Inductive Learning~\cite{2019_NLIL} uses transformer structure to get the non-chain-like rules which To extend the diversity of mined rules.

Whether it’s pure-symbolic or neural-symbolic method, none of these models consider the problem of utilizing the information about the head entity during the process of mining or using rules.

\subsection{Embedding-based Models}
A lot of previous works~\cite{TransE,ComplEx,ConvE,RotatE} concentrate on the embedding-based paradigm. Most of them design a scoring function to get a value for a triplet in the embedding space. Despite their simplicity, embedding-based models achieved good performance on reasoning. These works mostly rely on simple triplets, which means they ignore the environment of the entities and relations.

Some other embedding-based methods consider context information during inferring process. PTransE~\cite{ptranse} uses the path embedding from the source entity to target entity and the relation embedding to train the model jointly. \cite{chainofreasoning} further considers entities and entity type in the path. Except concentrating on the path, some methods~\cite{RGCN} utilize information from context with graph neural networks (GNN).
HittER~\cite{hitter} uses hierarchical Transformers to make contextualization based on a source entity’s neighborhood because they think the network architecture of GNNs is too shallow. 

These approaches prove contextual information which plays an important role in reasoning, and this aspires us to consider the environment of source entity in rule mining to some extent.
Although the performance on link prediction task is good, these works mostly are based on vector computation, so the symbolic meaning and interpretability are missing while rule mining methods perform better in this regard.

\section{Methodology}
\begin{figure*}[htb]
\centering
  \includegraphics[scale=0.75]{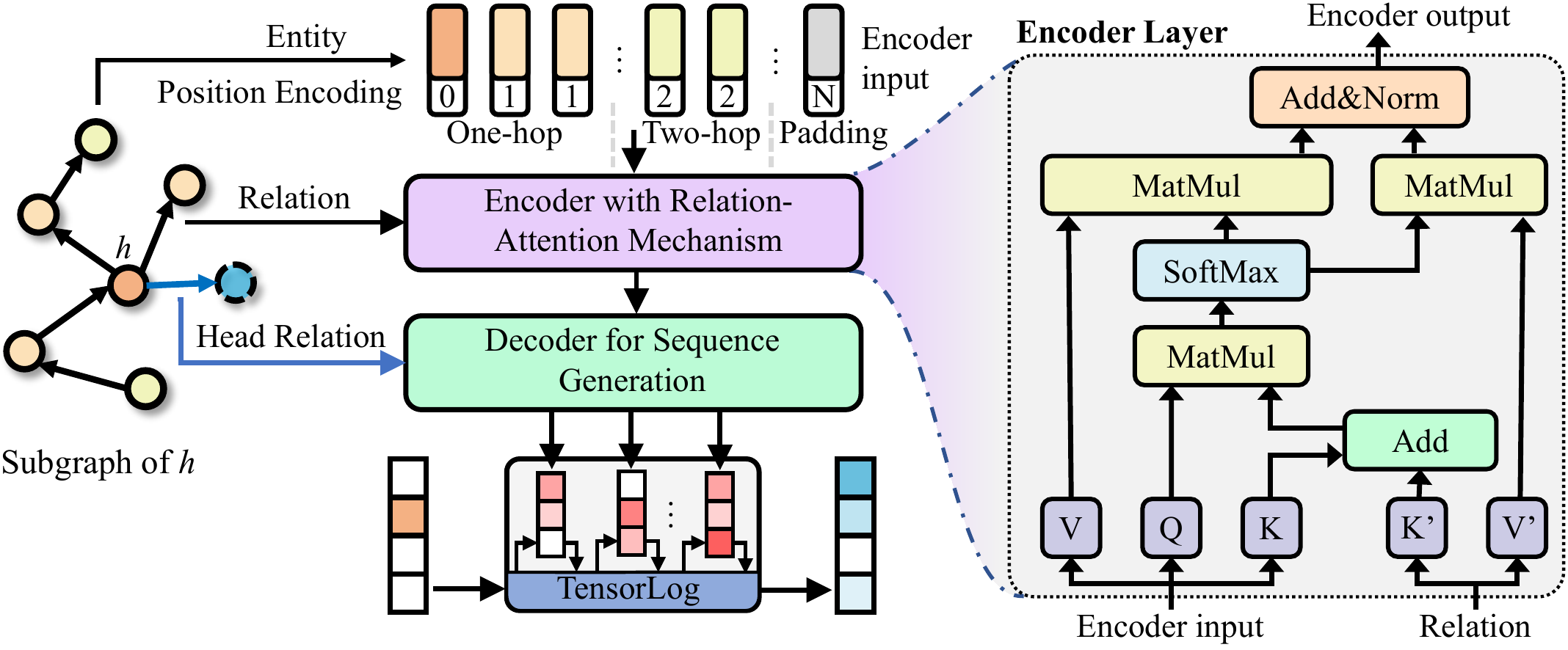}
  \caption{The framework of {\model} and details about relational attention mechanism.}
  \label{fig:model}
\end{figure*}

Knowledge Graph $\mathcal{G}$ is composed by a set of triplets like $\{(e_s, r, e_t) | r \in \mathcal{R}, e_s, \in \mathcal{E}, e_t \in \mathcal{E}\}$, where $\mathcal{E}$ is a countable set of entities and $\mathcal{R}$ is a set of relations, respectively.
For the task of rule mining, all the triplets that belong to $\mathcal{G}$ are given to the model, and for each head relation $r$, the model is supposed to find meaningful rules that are interpretable and understandable for humans.
The rule we want is in the following form:
\begin{equation}
    \label{rule:noconstant}
r\left( X, Y \right) \leftarrow r_1\left( X, Z_1 \right) \land ...\land r_T\left( Z_{T-1}, Y \right)
\nonumber
\end{equation}
where $T$ is the length of rule, $r$ and $r_i$ are relations belonging to $\mathcal{R}$, $X$, $Y$ and $Z_i$ are variables that can be replaced by specific entities and $r(X, Y)$ is a triplet.
The triplet on the left of the arrow is called \textit{head} of rule and the right part of the arrow is called rule \textit{body}.

In this section, to provide an intuition about each part of our model, we introduce the details about {\model}. Firstly, an encoder is designed for converting the subgraph structure to a sequence, so that the context information of head entity can be input to the Transformer framework with relational attention mechanism which is used to retain the structure information of the graph completely. 
Then we present how our model is deployed to utilize the output of encoder to support the rule mining process and generate the target sequence that represents rule body. Finally, we propose a case-based rule parsing algorithm to get symbolic rules from the parameters.

\subsection{\model}
Now we introduce the subgraph encoding process of the model, and Figure~\ref{fig:model} shows the details. 
The basic idea of our approach is to find  rules for the same head relation and let the model has the ability to choose the suitable rule when the contextual environment of the head entity is different.

For a query $(h, r, t)$, we assume that the subgraph around the head entity $h$ in KG contains the potential knowledge needed for understanding the environment, so we first extract the subgraph around $h$. More precisely, let $\mathcal{S}_k(h)$ be the subgraph which contains the $k$-hop (shortest distance to $h$) neighbor of $h$ in the KG and the set of edges connecting these entities.

As we stated before, KG is organized as a graph structure, while Transformer~\cite{transformer} is a seq2seq model, so we need to transfer the graph to a sequence. Specifically, the nodes in the subgraph are tokenized into a sequence, $S_{node} = [e_1, e_2, ...e_{num}, ...blank]$, where $num$ is the number of entities in the subgraph, and a special token \textit{blank} is needed for padding. As shown in the Figure~\ref{fig:model}, each node in the subgraph is extracted, and mapped to the initial embedding for entities. The shortest distance to the head entity $h$ is also added as position embedding.

Meanwhile, the distribution of entity occurrence is heavy tailed and hence it is hard to learn appropriate representations for each entity. To alleviate this problem, we consider using the type of relations to help represent the entity.
For each relation $r$, we define two randomly initialized embedding $r^{dom}$ and $r^{ran}$ representing the domain and range embedding of $r$, which can be learned during the training process. An entity $e$ gets its representation $x_e$ by addition of the type embedding of relations connecting to the entity and a randomly initialized embedding $y_{e}$ as follows:
\begin{equation}
\label{entity_embedding}
 x_e= \sum_{i=1}^{|\mathcal{R}|}b_{i}^{dom}r^{dom}_{i} + \sum_{i=1}^{|\mathcal{R}|}b_{i}^{ran}r^{ran}_{i} + y_{e}
\end{equation}
where $b_{i}^{dom}$ and $b_{i}^{ran}$ are the parameters which are determined by the numbers of different relation types after normalization, and can be given as:
\begin{equation}
\label{a}
 b_{i}^{dom} = \frac{n_{i}^{dom}}{\sum_{j=1}^{|\mathcal{R}|}n_{j}^{dom}}, \quad
 b_{i}^{ran} = \frac{n_{i}^{ran}}{\sum_{j=1}^{|\mathcal{R}|}n_{j}^{ran}}
\end{equation}
where $n_i^{dom}$ and $n_i^{ran}$ is the number of each type relation connecting to and connected from $e$, specifically.

Thus, the node sequence $S_{node}$ in mapped to a sequence of embedding $S_e = [x_1, x_2, ...x_{num}, ...blank]$.
Note that the above steps only use the entities in the subgraph without the edges between them. In order to utilize the edge information, we introduce relational attention mechanism. 
Different from the basic way of attention calculation in Transformer, we modify the steps to compute the attention $a_{ij}$ between entity $i$ and $j$ with relations connectingin the following way:
\begin{equation}
\label{attn_1}
a_{ij} = \frac{(x_iW^Q)(x_jW^K + \sum_{r=1}^{|\mathcal{R}|}k_rx_rW^{K'})}{\sqrt{d_k}}
\end{equation}
where $x_i$ and $x_j$ are the embedding of entity $i$ and $j$, respectively. $x_r$ is the embedding for relation $r$ which is different from $r^{dom}$ and $r^{ran}$. $W^Q$ and $W^K$ are the query and key matrices for entity, while $W^{K'}$ is the key matrix for relation. $d_k$ is the dimension of $W^K$. $k_r$ is defined as $k_r=1$ if $r(e_i, e_j) \in \mathcal{G}$, else $k_r=0$. 

Then the normalization step for attention is executed:
\begin{equation}
\label{attn_2}
 \alpha_{ij} = \frac{\exp a_{ij}}{\sum_{k=1}^{n}\exp a_{ik}}
\end{equation}

Similarly, we add the relation to the value computation step:
\begin{equation}
\label{attna_3}
 z_i = \sum_{j=1}^{n}\alpha_{ij}(x_jW^V + \sum_{r=1}^{|\mathcal{R}|}k_rx_rW^{V'})
\end{equation}
where $W^V$ is the value matrix for entity and $W^{V'}$ for relation. By doing this, the information of relations has been inserted into the encoding process.


The relational attention mechanism can ensure sufficient information exchange between each entity because the relations that clearly exist will be emphasized, while the relations that do not appear in the incomplete KG but may be correct will also be reflected because of the interaction process.
 
With the encoder output sequence $S_e'$, the decoder block does most of the lifting to aggregate the information together with the head relation. As we mentioned before, the rule mining process can be regarded as a sequence generation problem. For each step, the decoder generates the most suitable relation from $\mathcal{R}$, until the length of decoder output sequence $S_r$ reaches the rule length $T$.

Specifically, the rule sequence $S_r$ input to decoder starts with the head relations $r$'s embedding $x_r$, which means $S_r^0 = x_r$. After cross attention calculation with $S_e'$, the decoder gets a vector which implies the next relation. With this vector, an MLP function is deployed, and we can get the probability $\omega^i_t$ of relation $r_i$ in step $t$, which will be used in the reasoning process. The relation with the highest probability is chosen as the next relation added to $S_r$.
\begin{equation}
    \label{cross-attention}
    \omega_{t+1} = MLP(CrossAttention(S_r^{t}, S_e'))
\end{equation}
where $\omega_{t+1} \in \mathbb{R}^{|\mathcal{R}|\times 1}$ could be interpreted as the probabilities of all relations in step $t$. Let $r_{t+1}$ be the relation with the max probability, then $S_r^{t+1} = [S_r^{t}, x_{r_{t+1}}]$.

We repeat this step $T$ times, and get the complete rule body with length $T$. Moreover, considering that rule length can't be limited to a definite number $T$, we add a special relation \textit{self-loop} which connects entities with themselves, and we finally remove this relation from the rule body so that we can get rules with length less than $T$.

There is still a problem that it's different from the task of machine translation that we don't have labels to judge if the generated relation is an appropriate choice for each step. To overcome this issue, we deploy the Tensorlog framework here to get the prediction results like Neural-LP~\cite{yang2017differentiable} and DRUM~\cite{sadeghian2019drum} to indirectly supply supervisory signal. Each entity $e_i$ is represented as a one-hot vector $\mathbf{v}^{e_i} \in \{ 0, 1\}^{|\mathcal{E}|}$, and each relation $r_k$ is represented as an adjacent matrix $\mathbf{M}^{r_k}\in \left \{ 0,1 \right \}^{|\mathcal{E}| \times |\mathcal{E}|}$, where $\mathbf{M}^{r_k}_{ij}=1$ if $r_k(e_i, e_j) \in \mathcal{G}$, else $\mathbf{M}^{r_k}_{ij}=0$.

Via applying the entity vector and relation matrix, path queries could be answered by expanding path as follows:
\begin{equation}
    \mathbf{v}' = \mathbf{v}^{e_i} \mathbf{M}^{r_k}
\end{equation}
Note that $\mathbf{v}'$ is a multi-hot vector that refers to several entities, which means these entities are connected to $e_i$ via relation $r_k$.

For step $t$, the probability $\omega_t$ of all relations, which is generated in Equation~\ref{cross-attention}, are used by applying the above step in an indirect way as Equation~\ref{step}. Let $\mathbf{z}_{t}\in \mathbb{R}^{|\mathcal{E}|\times 1}$ be the vector representing the probability of all entities in step $t$, and it is a one-hot vector that represents the head entity if $t=0$. With $\mathbf{z}_{t-1}$ after $t-1$ step inference, $\mathbf{z}_{t}$ can be computed as follow:
\begin{equation}
\label{step}
 \mathbf{z}_{t}=\mathbf{z}_{t-1}\times \sum_{i=1}^{|\mathcal{R}|+1}\omega^{i}_{t}\mathbf{M}^{r_i}
\end{equation}
 The special relation $self-loop$ with an identity adjacency matrix with $\mathbf{M}^{r_{|\mathcal{R}|+1}} = I_{\mathcal{|E|}}$ is also considered. With this relation, the model can mine rule with length shorter than the length $T$.
 $\omega_{t}^{i}$ represents the probability of the relation $r_i$ as the relation in rules at step $t$.
 
 Finally, we get the $\mathbf{z}_{T}$, which is the final result after $T$ steps reasoning. For triplet $(h,r,t)$, the reasoning score is the similarity between the predicted vector $\mathbf{z}_T$ and the target entity vector $\mathbf{v}$:
\begin{align}
    \phi (t | h, r) = \mathbf{v} \cdot log [\mathbf{z}_T, \gamma]_+
\end{align}
where $[\mathbf{x}, \gamma]_+$ denotes the maximum value between each element of $\mathbf{x}$ and $\gamma$. The objective function {\model} is:
\begin{equation}
    \text{min} \;
     \left(-\sum_{(h,r,t) \in 
     \mathcal{G}}
     \phi {(t | h,r)}\right)
\end{equation}

\subsection{Rule Parsing}

 To decode symbolic rules from {\model}, we propose a rule parsing algorithm using the parameters learned from training process. The basic idea is to select appropriate relations with high weight.
With different triplets and the context information fed to the model, {\model} may output different parameters, so even for the same relation, the rules mined might be different.
Specifically, for a query $(h, r, t)$, we recover possible rules via parameters {$\alpha$}. In each step, we choose relations whose weights are over the threshold, and we will check if the entities in the previous step are linked with this relation. By performing this step cyclically until rule length reaches the max length $T$, the confidence of each rule is computed by multiplying the weights of relations selected.  output symbolic rules with high confidence. Finally, rules that may be useful in reasoning process for a triplet will be output.
The detailed procedure is shown in Algorithm \ref{alg:parser}.

Then we summarize the number occurrences of each rule, and for each time a rule is applied, a confidence score will be calculated with $\omega$ and the final score is the average confidence of this rule in different cases. Specifically, a rule set is defined, and for each triplet $(h, r, t)$, we apply Algorithm~\ref{alg:parser} to parse rules from it. Finally, the score will be calculated for all the rules in $R$.
\begin{algorithm}[H]
    \caption{Decode symbolic logical rules from model}
    \label{alg:parser}
    \textbf{Input}: attention $\{\omega_t | t=1,2...T\}$ for each triplet\\
    \textbf{Initialize}: $P=\{([P_r], [P_e], w)\}$ , $P_r=\emptyset$, $P_e=$ head entity, $w$ represents confidence;\\
    \textbf{Output}: $R$
    \begin{algorithmic}[1]
        \FOR{$t = 1:T$}
            \STATE \textit{// Scale the attention}
            \STATE $\omega_t = \omega_t / max(\omega_t)$
            \FOR{$([P_r, P_e], w) \in P$}
                \FOR{$\omega_t^{r_p} \in \omega_t > thr$}
                    \STATE \textit{// Expand a new path if possible}
                    \FOR {$n \in \mathcal{E}$ can be linked with $P_e[-1]$ via $r_p$}
                        \STATE \textit{// Compute the new confidence}
                        \STATE$w' = w \times \omega_t^{r_p}$
			            
                        \STATE add $([P_r+{r_p}], [P_e+n], w')$ to $P$
                    \ENDFOR
                \ENDFOR
                \STATE \textit{// Remove the old path}
                \STATE remove $([P_r, P_e], w)$ from $P$
            \ENDFOR
        \ENDFOR
        \FOR{$([P_r, P_e], w) \in P$}
            \STATE \textit{// $R[r, p_r]$ is a list stores confidence scores }\\
            \STATE  add $w$ to $R[r, p_r]$\\
        \ENDFOR
    \end{algorithmic}
    
\end{algorithm}

\begin{table*}[t]
\centering
\resizebox{\textwidth}{!}{ 
\begin{tabular}{ccccccccccccc} 
\toprule
\multirow{3}{*}{\textbf{Methods}} & \multicolumn{4}{c}{\textbf{UMLS}}                                             & \multicolumn{4}{c}{\textbf{FB15K-237}}                                        & \multicolumn{4}{c}{\textbf{WN18RR}}                                            \\ 
\cline{2-13}
                                  & \multirow{2}{*}{\textbf{MRR}} & \multicolumn{3}{c}{\textbf{HIT}}              & \multirow{2}{*}{\textbf{MRR}} & \multicolumn{3}{c}{\textbf{HIT}}              & \multirow{2}{*}{\textbf{MRR}} & \multicolumn{3}{c}{\textbf{HIT}}               \\ 
\cline{3-5}\cline{7-9}\cline{11-13}
                                  &                               & \textbf{@1}   & \textbf{@3}   & \textbf{@10}  &                               & \textbf{@1}   & \textbf{@3}   & \textbf{@10}  &                               & \textbf{@1}   & \textbf{@3}   & \textbf{@10}   \\ 
\midrule
TransE                            & .668                          & 46.8          & 84.5          & 93.0          & .294                          & -             & -             & 46.5          & .226                          & -             & -             & 50.1           \\
DistMult                          & .753                          & 65.1          & 82.1          & 93.0          & .241                          & 15.5          & 26.3          & 41.9          & .430                          & 39.0          & 44.0          & 49.0           \\
ComplEx                           & .829                          & 74.8          & 89.7          & 96.1          & .247                          & 15.8          & 27.5          & 42.8          & .440                          & 41.0          & 46.0          & 51.0           \\
ComplEx-N3                        & -                             & -             & -             & -             & .370                          & -             & -             & \textbf{56.0} & \textbf{.480}                 & -             & -             & 57.0           \\
ConvE                             & .908                          & 86.2          & 94.4          & 98.1          & .325                          & 23.7          & 35.6          & 50.1          & .430                          & 40.0          & 44.0          & 52.0           \\
TuckER                            & -                             & -             & -             & -             & \textbf{.358}                 & \textbf{26.6} & \textbf{39.4} & 54.4          & .470                          & \textbf{44.3} & 48.2          & 52.6           \\
RotatE                            & \textbf{.948}                 & \textbf{91.4} & \textbf{98.0} & \textbf{99.4} & .338                          & 24.1          & 37.5          & 53.3          & .476                          & 42.8          & \textbf{49.2} & \textbf{57.1}  \\ 
\midrule
PathRank                          & .197                          & 14.7          & 25.6          & 37.6          & .087                          & 7.4           & 9.2           & 11.2          & .189                          & 17.1          & 20.0          & 22.5           \\
Neural-LP(T=2)*                   & .751                          & 63.0          & 84.7          & 94.0          & .189                          & 12.7          & 20.6          & 31.3          & .371                          & 35.9          & 37.4          & 39.6           \\
Neural-LP(T=3)*                   & .735                          & 62.7          & 82.0          & 92.3          & .239                          & 16.0          & 26.1          & 39.9          & .425                          & 39.4          & 43.2          & 49.2           \\
DRUM(T=2)*                        & .791                          & 64.5          & 92.7          & 96.8          & .225                          & 17.1          & 25.4          & 35.8          & .379                          & 36.8          & 38.5          & 40.9           \\
DRUM(T=3)*                        & .784                          & 64.3          & 91.2          & 97.2          & .328                          & 24.7          & 36.2          & 49.9          & .441                          & 41.2          & 45.6          & 51.6           \\
M-Walk                            & -                             & -             & -             & -             & .232                          & 16.5          & 24.3          & -             & .437                          & 41.4          & 44.5          & -              \\
\textbf{{\model}(T=2)}                    & .851                          & 73.6          & \textbf{96.6} & \textbf{98.8} & .237                          & 17.4          & 25.7          & 36.0          & .381                          & 36.6          & 38.8          & 41.1           \\
\textbf{{\model}(T=3)}                    & \textbf{.857}                 & \textbf{75.2} & 95.8          & 98.4          & \textbf{.342}                 & \textbf{25.5} & \textbf{37.4} & \textbf{51.3} & \textbf{.452}                 & \textbf{41.7} & \textbf{46.5} & \textbf{53.0}  \\
\bottomrule
\end{tabular}
}

\caption{Link prediction results on dataset \textit{UMLS}, \textit{FB15K-237} and \textit{WN18RR}. Note that for some algorithm, several entities may get the same score. Instead of computing the rank of the right answer as $m+1$ where $m$ is the number of entities with higher possibilities, we select a random rank for the right answer among the entities with the same possibility. Some methods reported their results in the previous setup in their original paper, and we rerun these methods with the same evaluation process in our way. The results with [*] is reported with our evaluation protocol.}
\label{tab:link_prediction}
\end{table*}

\section{Experiment}

\subsection{Dataset}
We conduct experiments on three different datasets which are introduced as follows, and Table \ref{tab:datasets_statistics} summarizes the data statistics. We count the average degree for each dataset, because the sparsity has an impact on the choice of subgraph.

\begin{itemize}
\item \textit{UMLS} \cite{umls}: Unified Medical Language System, is a knowledge graph
that brings together many health and biomedical vocabulary and standards.
\item \textit{FB15K-237} \cite{fb15k237}: This dataset is a subset from Freebase\cite{bollacker2008freebase} and removes the inverse relation. It stores commonsense facts such as topics in movies, actors, awards, etc.

\item \textit{WN18RR} \cite{ConvE}: WN18RR is a link prediction dataset created from WN18, which is a subset of WordNet. It's created to ensure that the evaluation dataset does not have inverse relation test leakage.
\end{itemize}

\begin{table}[htb]
\centering
\begin{tabular}{lrrr}
\toprule
\textbf{Datasst}      & \textbf{UMLS} & \textbf{FB15K-237} & \textbf{WN18RR} \\ \hline
\textbf{\#Triplet}    & 5,960         & 310,116            & 93,003          \\
\textbf{\#Entity}     & 135           & 14,541             & 40,943          \\
\textbf{\#Relation}   & 46            & 237                & 11              \\
\textbf{\#Avg.deg} & 88.3          & 42.7               & 4.5             \\ \toprule
\end{tabular}
\caption{Knowledge graph datasets statistics.}
\label{tab:datasets_statistics}
\end{table}

\subsection{Experiment Setup}
The experiments are implemented with Pytorch framework and are trained on RTX3090 GPU. ADAM optimizer is used for parameter tuning and the learning rate is set to 0.0001. 
Our model consists of the encoder and decoder block, which are two-layer Transformer and each layer has 
$6$
heads by default. The dimension of entity 
is set to 200, as well as the position encoding dimension. We also try 6 heads and 3 layers in Transformer and larger dimension, which result in a little difference. Dropout is applied with a possibility $p=0.1$. The $\gamma$ used as threshold is set to be $10^{-20}$.

Since the average degree is different for each dataset, to avoid the input sequence of entities being too long, the choice of subgraph is different, too. 
We extract one-hop subgraph for \textit{FB15K-237}, two-hop subgraph for \textit{WN18RR}. For \textit{UMLS}, the shortest distance is the same as the rule length.
Meanwhile, the maximum number of entities which is based on the sparsity of each dataset is different. 
Considering that relations with a large number of neighbors , like \textit{hasGender}, don't contain much information relative to the huge number of entities, a max number of direct neighbor linked by one relation is set for each entity.
For \textit{UMLS}, we set the max number of context entities and max number of neighbors for each entity as 140 and 40 respectively. For \textit{FB15K-237} and  \textit{WN18RR}, they are 70 and 40, and 40 and 10 respectively.
If the number of context or neighbors exceeds our settings, we randomly select the same number of entities as the setting.

\subsection{Link Prediction}
For the link prediction task, each triplet $(h, r, t)$ in test dataset and  its inverse triplet $(t, r^{-1}, h)$ are given to the model with the tail entity  masked, and the goal is to predict the masked tail entity among all entities. 
Each candidate entity in KG will get a score according to the inference result, which is used to sort them, and the ground truth entities are filtered out of the ranking list.
Here, we adopt the evaluation metrics the same as previous works~\cite{TransE}, Hit@1, Hit@3, Hit@10, and mean reciprocal rank(MRR).

We compare our model with some embedding methods including TransE~\cite{TransE}, DisMult~\cite{DistMult}, ConvE~\cite{ConvE}, TuckER~\cite{balavzevic2019tucker} and RotatE~\cite{RotatE}, and rule mining methods like PathRank~\cite{pathrank}, Neural-LP~\cite{yang2017differentiable}, DRUM~\cite{sadeghian2019drum} and M-Walk~\cite{mwalk}. Compared to embedding methods, rule mining methods can provide the interpretability, which is an advantage to pure embedding methods. The results are shown in the Table~\ref{tab:link_prediction}.

As the Table shows, our model outperforms the other rule-based approaches on three datasets. On \textit{UMLS}, the competitive results demonstrates the effectiveness of rule mining approaches. {\model} offers an absolute improvement in hit@1 about 10.9\% (relatively 16.9\%) compared against baseline on \textit{UMLS}. On dataset \textit{FB15K-237} and \textit{WN18RR}, our model also gets the best results on rule mining methods which improves the overall results about 1.4\% and 1.1\% (relatively 4.3\% and 2.3\%), respectively. 
{\model} achieves better performance compared to other rule mining proves the effectiveness of our method, and we think this improvement is because other rule-based methods don't consider the head entity and its subgraph and confirms the correctness of our hypothesis.

\subsection{Quality Assessment of Minded Rules}
In order to have an objective evaluation of the mined rules, we adopt the Standard Confidence (SC)~\cite{galarraga2013amie} to assess the rules mined by different methods. Specifically, the average score of rules is calculated to show the quality. Given a list of rules which is ranked by their confidence calculated by their algorithm that is different from SC, we report the average SC of the top$K$ ($K=50, 100, 200, 500$) rules.
The rules of Neural-LP and DRUM are provided by their original code. The 
Standard Confidence $SC(\mathbf{B}\rightarrow r(X, Y))$ is calculated as follows:
\begin{equation}
\nonumber
     \frac{\#(X, Y):\exists Z_1...Z_m : \mathbf{B} \wedge r(X, Y)}{\#(X, Y): \exists Z_1,\dots,Z_m : \mathbf{B}}
\end{equation}
where \textbf{B} represents the rule body, $X$, $Y$ and $Z$ are entity variables, and $r$ is the head relation. This score regards facts that are not in KG as false, in another 
word,
it implements a closed world setting.

Table~\ref{tab:confidence} reports the results. 
We compare {\model} with other differentiable rule mining methods. The main difference between these methods and ours is that {\model} could parse the rule according to the environment, but others parse the same rules for each triplet with the same relation.
There is a significant improvement in standard confidence with our model and algorithm. Specifically, the standard confidence on \textit{UMLS} improves about 43.7\% and on \textit{FB15K-237}, the improvement is 25.5\%. This is not only because the model generates suitable rules during training, but also because our parsing algorithm outputs the rules which can be mapped to an existing path in KG.

\begin{table}[h]
\centering
\small
\setlength{\tabcolsep}{1.5mm}{

\begin{tabular}{ccccccc} 
\toprule
\multirow{3}{*}{\textbf{Methods}} & \multicolumn{3}{c}{\textbf{UMLS}}              & \multicolumn{3}{c}{\textbf{FB15K-237}}          \\ 
\cline{2-7}
                                  & \multicolumn{3}{c}{\textbf{TOP }}              & \multicolumn{3}{c}{\textbf{TOP }}               \\ 
\cline{2-7}
                                  & \textbf{50}   & \textbf{100 } & \textbf{200 }  & \textbf{50}   & \textbf{200 } & \textbf{500 }   \\ 
\hline
Neural-LP(T=2)             & .228          & .239          & .221           & .020          & .044          & .033            \\
Neural-LP(T=2)             & .104          & .145          & .153           & .020          & .031          & .034            \\ 
\hline
DRUM(T=2)                  & .400          & .350          & .303           & .058          & .036          & .048            \\
DRUM(T=3)                  & .340          & .284          & .202           & .020          & .039          & .027            \\ 
\hline
{\model}(T=2)    & \textbf{.837} & \textbf{.793} & \textbf{.740 } & .241          & \textbf{.338} & \textbf{.310 }  \\
{\model}(T=3)     & .680          & .652          & .573           & \textbf{.313} & .322          & .282            \\
\toprule
\end{tabular}
}
\caption{Average confidence of top ranked rules on datasets \textit{UMLS} and \textit{FB15K-237}. The superscript [2] or [3] means with rule length 2 or 3, respectively.}
\label{tab:confidence}
\end{table}

\subsection{Case Study}
As we introduced, for different prediction tasks,
we hope our model generate suitable rules in a better way with context of head entity into consideration rather than using them in the same order.

\begin{figure}[tb]
  \centering
  \includegraphics[scale=0.47]{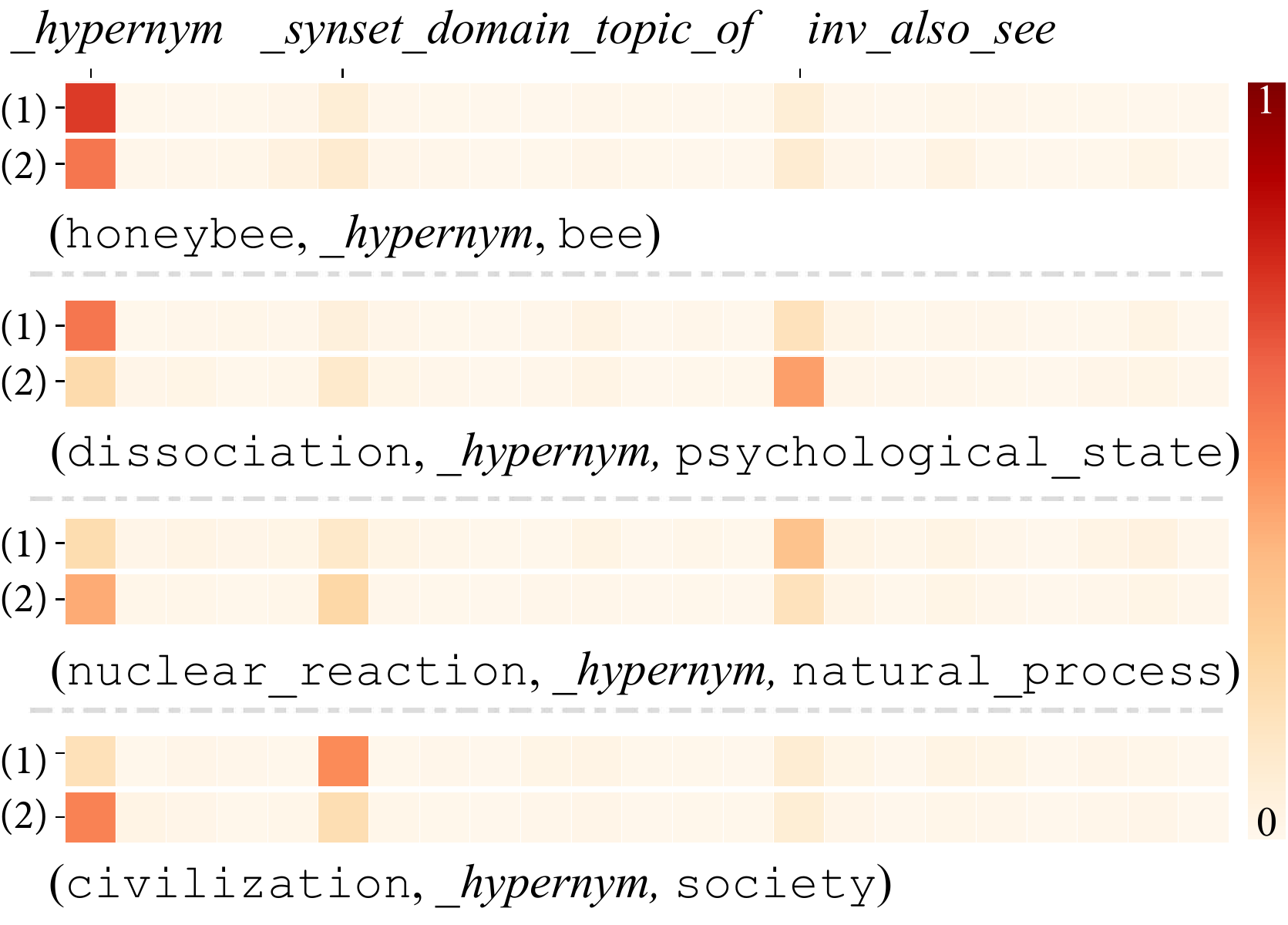}
  \caption{Decode output of four different triplets with the same head relation \textit{\_hypernym}. (1) and (2) means the first and second step.}
  \label{fig:dec_outputs}
\end{figure}

To verify the ability of generating appropriate rules of our model, we choose four test triplets from \textit{WN18RR} that are shown in Figure \ref{fig:dec_outputs}. 

Input these triplets into our model with rule length $T=2$, and the output contains four sets of data which represent the probability of each relation for two steps in different cases.
The heatmap in Figure~\ref{fig:dec_outputs} shows the output.
Each row in the figure represents the probability for all relations in a step, and each pair of rows correspond to rules with different scores for the specific triplet which is showed below each subfigure.
The darker the color is, the higher probability the relation has for current step.
Note that we added inverse relations and \textit{self-loop} so there are more than twice relations in the dataset. The relations with high probability are marked in the top of the figure.

As the figure shows, the four triplets have the same head relation \texttt{hypernym} which means a word that is more generic than a given word, while the head entities are different. 
for different triples with the same relation to be predicted, the probability of relations generated for reasoning in each step are not the same.
Take the first case as an example, the rule with the highest confidence is '\texttt{hypernym(X, Z)} $\leftarrow$ \texttt{hypernym(X, Y)} $\wedge$ \texttt{hypernym(Y, Z)}'. While for the second example, the rule is '\texttt{hypernym(X, Z)} $\leftarrow$ \texttt{hypernym(X, Y)} $\wedge$ \texttt{inv\_alsoSee(Y, Z)}' where \texttt{inv} means inverse relation.
These different but all correct rules shows that our method successfully generate different rules and contextual information of triplets do have an impact on rule reasoning.

\section{Conclusion}

In this paper, we draw attention to using context to assist rule mining.
We regard it as a sequence generation problem, and design a converter turning graph structure into a sequence.
We propose a transformer-based model, {\model}, which utilizes the subgraph of head entity when learning rules over knowledge graph.  
The experiment results show that in a specific environment, our proposed model can mine and select different suitable rules. The performance on link prediction task and rule parsing also improves with our model and parsing algorithm.
Future work may focus on a more effective way of rule mining and consider more complex forms of rules.

\section*{Acknowledgements}
This work is funded by NSFCU19B2027/91846204. 

\appendix

\bibliography{anthology,custom}
\bibliographystyle{acl_natbib}

\end{document}